\documentclass{aastex62}

\usepackage{graphicx}
\usepackage{enumerate}
\usepackage{url}
\usepackage{xcolor}
\usepackage{upgreek}
\usepackage{longtable}
\usepackage{mathrsfs}
\usepackage{kellymacros2}
\usepackage{natbib}
\usepackage{amsmath}

\newcommand{\RomanNumeralCaps}[1]
    {\MakeUppercase{\romannumeral #1}}

\shorttitle{SDSS HET \kep\ APOGEE Benchmark Binaries}
\shortauthors{Hambleton et al.}

\begin{document}

\title{The SDSS-HET Survey of \kep\ Eclipsing Binaries. A Sample of Four Benchmark Binaries.}

\author[0000-0001-5473-856X]{Kelly Hambleton}
\affil{Department of Astronomy \& Astrophysics, Villanova University, 800 East Lancaster Ave., Villanova, PA 18085}
\email{kelly.hambleton@villanova.edu}

\author[0000-0002-1913-0281]{Andrej Pr\v{s}a}
\affil{Department of Astronomy \& Astrophysics, Villanova University, 800 East Lancaster Ave., Villanova, PA 18085}

\author[0000-0003-0556-027X]{Scott W. Fleming}
\affil{Space Telescope Science Institute, 3700 San Martin Dr, Baltimore, MD 21218}

\author[0000-0001-9596-7983]{Suvrath Mahadevan}
\affil{Department of Astronomy \& Astrophysics,  The Pennsylvania State University, 525 Davey Lab, University Park, PA-16802}
\affil{Center for Exoplanets $\&$ Habitable Worlds, The Pennsylvania State University, 525 Davey Lab, University Park, PA-16802}

\author[0000-0003-4384-7220]{Chad F. Bender}
\affil{Department of Astronomy and Steward Observatory, University of Arizona, Tucson, AZ 85721}

\begin{abstract}
    The purpose of this work is to extend a sample of accurately modeled, benchmark-grade eclipsing binaries with accurately determined masses and radii. We select four ``well-behaved'' \kep\ binaries, KIC\,2306740, KIC\,4076952, KIC\,5193386 and KIC\,5288543, each with at least 8 double-lined spectra from the \apogee\ instrument that is part of the Sloan Digital Sky Surveys \RomanNumeralCaps{3} and \RomanNumeralCaps{4}, and from the Hobby-Eberly High Resolution Spectrograph. We obtain masses and radii with uncertainties of 2.5\% or less for all four systems. Three of these systems have orbital periods longer than 9 days, and thus populate an under-sampled region of the parameter space for extremely well-characterized detached eclipsing binaries. We compare the derived masses and radii against \mesa\ \mist\ isochrones to determine the ages of the systems. All systems were found to be coeval, showing that the results are consistent across \mesa\ \mist\ and \ph.  
\end{abstract}


\section{Introduction}

To derive accurate stellar masses and radii, systems with strong constraints are required. When considering multiple star systems, the optimal case is a system of three or more stars where every component is eclipsing, cf.~\citet{Carter2011a}, as this places stringent constraints on the system geometry. However, these objects are rare and the software required to accurately model all time-dependent aspects of these systems within a Roche framework is still under construction \citep{Conroy2020}. Another way to obtain accurate parameters is to look for totally eclipsing binaries (EBs) as the constraints for those cases, while not as stringent as for triples, provide us with better handles than partially eclipsing EBs. The goal of this work is to expand the library of accurate masses and radii upon which stellar evolutionary models can be validated.

Of course, alongside each well-behaved totally-eclipsing binary system, precise data are required to obtain the parameter accuracy goals. The \kep\ satellite \citep{Borucki2010} has provided a treasure trove of objects with highly precise data, which include 2922 binary stars \citep{Prsa2011,Slawson2011,Kirk2015}. The level of detail obtained by the \kep\ satellite has enabled approximate mass estimates to be extracted from \kep\ light curves without the need for radial velocities through modeling ellipsoidal variations and Doppler boosting (see \eg\ \citealt{Carter2011b}). This method offers the possibility of obtaining masses for objects that are not good candidates for radial velocity followup to a precision of $\sim$10\%. For the purposes of calibrating stellar evolutionary models, however, we require masses and radii to a precision of $\le$3\% \citep{Morales2010} and aim for $\le$2\%. For this reason this work is based on a selection of four well-behaved eclipsing binary stars with a significant number of well-sampled radial velocity data points. 

For the definition of well-behaved or benchmark-grade binary systems, we adopt the criteria of \citet{Torres2010}, based on that of \citet{Andersen1991}, which states: the stellar components must appear to have evolved as if they were single stars, which excludes any mass transfer; the spectra must be of high resolution and signal-to-noise; the period should be easily determined, \ie\ there is no measurable apsidal motion; and the eclipses must both be significant such that the ratio of the temperatures can be determined through direct measurement (it is worthy of note that the absolute temperature of one of the components, though spectral fitting or similar, is necessary to remove degeneracies from the model). With these stringent requirements, we are able to provide a discriminating test of theoretical evolution models \citep{Southworth2004}. A catalog of established benchmark-grade eclipsing binaries that are subject to these criteria can be found in the DEBCat Library\footnote{\url{http://www.astro.keele.ac.uk/jkt/debcat/}} \citep{Southworth2014}.

This paper is the third in a series of papers describing objects from the {\emph{SDSS-HET Survey of Kepler EBs}} \citep{Bender2012,Mahadevan2019}, which focus on the \kep\ data and radial velocity data from the Sloan Digital Sky Survey (SDSS) and the Hobby-Eberly Telescope (HET). Throughout this paper, we will refer to details described in the original paper, \citet{Mahadevan2019}, hereafter M19. Other works that have taken advantage of the overlap between \apogee\ and \kep\ to determine precise solutions for eclipsing binaries include \citet{Mahadevan2019}, \citet{Cunningham2019} and \citet{Gaulme2016}. The purpose of this paper is to provide benchmark-quality binary star models of four objects where the derived masses and radii are precise to better than 2.5\% or better. 

In Section\,\,\ref{sec:facilities} we describe the facilities and data products used for this work. In Section\,\ref{sec:analysis} we discuss the detailed analysis of the data, including Markov chain Monte Carlo methods and Gaussian process application for the noise model. In Section\,\ref{sec:benchmark} we describe the four benchmark-grade binary stars that have been analysed, and in Section\,\ref{sec:discuss} we summarize the outlined work and provide concluding remarks.

\section{Facilities and Datasets\label{sec:facilities}} 

\subsection{The \kep\ Mission\label{sec:kepler}}

The \kep\ satellite \citep{Borucki2010}, launched in 2009, simultaneously and continuously observed $\sim$156,000 stars for $\sim$4\,yr. The telescope had a 0.95\,m primary mirror and a high-precision white-light photometer with 42 CCDs. During its primary mission it monitored a 115 square-degree region of the sky in the direction of Cygnus and Lyra in an Earth trailing, heliocentric orbit. The nearly continuous observations generated unprecedented light curves that paved the way for numerous scientific advances in planetary science - the primary mission objective - but also for auxiliary goals such as binary star science. 

For the purpose of this work we used \kep\ long cadence observations ($\sim$30\,min exposure time), which were downloaded from MAST (Mikulski Archive for Space Telescope). We elected to use the simple aperture photometry, which has been minimally processed by the \kep\ pipeline. Prior to performing binary star analysis, we detrended and normalized the light curves by fitting a low order (order\,=\,1--4) Legendre polynomial to each segment of data between breaks. Obvious outliers were additionally removed by eye. 

As the \kep\ light curve uncertainties are known to be underestimated \citep{Bryson2011}, we independently determined the uncertainties for each object by finding the standard deviation of multiple segments of the light curve. To do this, each out-of-eclipse segment was detrended by applying a high order (order\,=\,3--20) polynomial. We aggressively detrended the lightcurve to remove all instrumental systematics and deviations due to spots and other astrophysical processes for the purpose of error estimation only . For each object, we took the median average of the standard deviation for 10 segments to determine the final uncertainty. To ensure the uncertainties were properly estimated, we additionally included a Gaussian noise term in our fitting procedure, which is discussed in detail in \S\ref{sec:GP}. 

\subsection{The SDSS-III APOGEE Spectrometer\label{sec:apogee}}

The Apache Point Observatory Galactic Evolution Experiment (APOGEE, \citealt{Majewski2017}) is a fiber-fed, multi-object spectrometer on the 2.5\,m telescope of the Sloan Digital Sky Survey (SDSS \citealt{York2000,Gunn2006}) located at Apache Point Observatory. It is a near-infrared instrument with a wavelength range of 1.51 to 1.68\,$\mu$m and a resolution of $\lambda/\Delta\lambda \sim 22,500$. APOGEE can simultaneously observe 300 objects over a $3\deg$ diameter field of view. 

The spectra discussed herein were observed during 2011 and became publicly available with Data Release 10 \citep{Ahn2013}. We start with the ``apVisit'' spectra produced by the SDSS pipeline, which are combined exposures that typically amount to a little over one hour of total observation time. We performed additional post-processing prior to generating radial velocities. We applied a low order polynomial to remove continuum and normalize each spectrum. We additionally corrected for the imperfect correction of telluric absorption lines by manually interpolating over neighboring pixels. For additional details, we refer the reader to M19.

\subsection{The HET High-Resolution Spectrograph\label{sec:hrs}}

To complement the APOGEE spectroscopic data, we obtained optical spectra from the High-Resolution Spectrograph (hereafter HRS) on the Hobby-Eberly Telescope located on the McDonald Observatory. HRS is a visible-light, fiber-fed, cross-dispersed echelle spectrometer with a wavelength range of 4076 to 7838\,\AA\ and a resolution of $\lambda/\Delta\lambda \sim 30,000$.

Calibrations, including biases, flats and ThAr wavelength references, were obtained at the end of each night and additionally either before or after each observation. The HRS spectra were then channeled through a semi-automated pipeline which applies image processing, spectral extraction and wavelength calibration. As with the APOGEE spectra, continuum normalizing was performed as part of post-processing. We selected segments of the spectra that contain telluric contamination of 0.5\% or less. Eight segments of spectra were used as part of our analysis: 4390--5025\,\AA, 5100--5410\,\AA, 5475--5680\,\AA, 5770--5855\,\AA, 6020--6260\,\AA, 6365--6430\,\AA, 6620--6850\,\AA\ and 7450--7580\,\AA. The spectra in our sample were observed between 2011 and 2013. Additional information regarding the HET reduction and pipeline can be found in the Appendix within M19.

\section{Analysis Techniques\label{sec:analysis}}

\subsection{Measurement of Radial Velocities \label{rvanalysis}}

A requirement for our sample of benchmark-grade binary systems was that the spectra contained signatures of both stellar components. With both stars visible in the spectra, the mass ratio and the size of the orbit can be determined, which is imperative for obtaining accurate masses and radii.

To obtain radial velocities from the observed spectra, we applied our SXCORR implementation of TODCOR (Two Dimensional Cross Correlation, \citealt{Zucker2003}), which allows for segments of a spectrum with variable lengths. Previous works contain extensive descriptions of SXCORR, \ie\ \citet{Bender2008,Bender2012,Lockwood2014} and we refer the interested reader to the descriptions therein.

For the APOGEE spectra, we used synthetic templates generated from the PHOENIX-based BT-Settl model grid \citep{Allard2011}. The templates used for the HRS spectra consist of observations of early F through mid M dwarfs that were observed concurrently with the HRS EB observations, and additionally BT-Settl templates to ensure the full range of $T_\mathrm{eff}$, [M/H] and $\log g$ is covered. To generate a template from the BT-Settl library, we convolved the raw synthetic spectrum to the proper resolution (22,500 for APOGEE and 30,000 for HRS) and re-sampled to three pixels per resolution element. We additionally applied rotational broadening kernels based on the appropriate stellar parameters \citep{Gray1999,Claret2012}.  See Table\,\ref{tab:rvs} for the radial velocity measurements.

\begin{table}
\caption{HRS $\&$ APOGEE RV Measurements. \label{tab:rvs}}	
\begin{center}
\begin{tabular}{ccccc}
\hline
{\bf{UT Date}}    & {\bf{BJD-2,400,000}} & {$\bold{V_{A}}$}   & {$\bold{V_{B}}$}& {\bf{Instrument}}  \\
&                             & (km s$^{-1}$)    & (km s$^{-1}$) & \\
\hline
\multicolumn{5}{l}{\emph{KIC\,2306740}}\\
\hline
2011 Jul 21 &55763.657622	&	-47.095$\pm$0.098&	85.573$\pm$0.188 & HRS\\
2011 Aug 24	&55797.798841	&	80.442$\pm$0.094&	-58.287$\pm$0.174 & HRS\\
2011 Aug 28	&55801.772678	&	6.212$\pm$0.098& 	25.530$\pm$0.194 & HRS\\
2011 Oct 06	&55840.667652	&	57.174$\pm$0.100&	-32.382$\pm$0.188 & HRS\\
2011 Oct 10	&55844.647447	&	-27.296$\pm$0.114&	63.953$\pm$0.210 & HRS\\
2011 Oct 26	&55860.612849	&	70.338$\pm$0.244&	-47.579$\pm$0.460 & HRS\\
2011 Sep 07	&55811.613101	&	14.143$\pm$0.533&	14.403$\pm$0.847 & APG\\
2011 Oct 06	&55840.593371	&	58.339$\pm$0.468&	-34.714$\pm$0.747 & APG\\
2011 Oct 17	&55851.578556	&	43.858$\pm$0.476&	-19.668$\pm$0.849 & APG\\
\hline
\multicolumn{5}{l}{\emph{KIC\,4076952}}\\
\hline
2011 Sep 01	&	2455805.79105	&	35.382$\pm$0.500	&	-4.510$\pm$0.375	&	HRS	\\
2011 Sep 02	&	2455806.801811	&	-6.201$\pm$0.605	&	52.138$\pm$0.407	&	HRS	\\
2011 Sep 14	&	2455818.757964	&	-39.184$\pm$0.988	&	97.335$\pm$0.884	&	HRS	\\
2011 Sep 29	&	2455833.72453	&	76.907$\pm$0.48	    &	-60.358$\pm$0.372	&	HRS	\\
2011Oct 24	&	2455858.639961	&	-24.586$\pm$0.64	&	75.868$\pm$0.360	&	HRS	\\
2011 Nov 16	&	2455881.586275	&	79.316$\pm$0.307	&	-64.242$\pm$0.428	&	HRS	\\
2011 Sep 19	&	2455823.72736952&   79.902$\pm$0.336	&	-63.767$\pm$1.268	&	APG	\\
2011 Oct 17	&	2455851.6496057	&	67.395$\pm$1.16 	&	-47.523$\pm$3.203	&	APG	\\
\hline
\multicolumn{5}{l}{\emph{KIC\,5193386}}\\
\hline
2011 May 09	&	2455690.841836	&	7.638$\pm$0.98	    &	-97.698$\pm$0.392	&	HRS	\\
2011 Jun 24	&	2455736.715402	&	-9.475$\pm$0.28	    &	-76.520$\pm$0.490	&	HRS	\\
2011 Jul 06	&	2455748.705552	&	-54.475$\pm$0.86	&	-21.673$\pm$0.332	&	HRS	\\
2011 Jul 19	&	2455761.661753	&	-55.979$\pm$0.42	&	-20.181$\pm$0.360	&	HRS	\\
2011 Aug 19	&	2455792.795279	&	-35.731$\pm$0.400	&		    -           &	HRS	\\
2011 Aug 28	&	2455801.787839	&	-20.266$\pm$0.56	&	-62.673$\pm$0.318	&	HRS	\\
2011 Sep 09	&	2455813.703058	&	-43.735$\pm$0.734	&		    -           &	APG	\\
2011 Sep 19	&	2455823.727087	&	-29.297$\pm$0.713	&	        -       	&	APG	\\
2011 Oct 06	&	2455840.661733	&	7.215$\pm$0.586	    &		    -           &	APG	\\
2011 Oct 15	&	2455849.578958	&	-81.657$\pm$0.855	&	10.365$\pm$2.620	&	APG	\\
2011 Oct 17	&	2455851.649347	&	-87.364$\pm$0.998	&	13.654$\pm$2.999	&	APG	\\
2011 Nov 01	&	2455866.569977	&	-30.531$\pm$0.914	&		                &	APG	\\
\hline
\multicolumn{5}{l}{\emph{KIC\,5288543}}\\
\hline
2011 Sep 09	&	56151.614362	&	-45.535$\pm$0.210	&	129.722$\pm$0.417	&	HRS	\\
2011 Sep 19	&	56154.83568	&	-49.213$\pm$0.204	&	133.655$\pm$0.405	&	HRS	\\
2011 Oct 06	&	56204.697696	&	94.125$\pm$0.213	&	-55.371$\pm$0.424	&	HRS	\\
2011 Oct 15	&	56209.679998	&	-15.013$\pm$0.204	&	89.170$\pm$0.378	&	HRS	\\
2011 Oct 17	&	56213.690411	&	-49.740$\pm$0.343	&	135.083$\pm$0.527	&	HRS	\\
2011 Nov 01	&	56262.546422	&	-21.577$\pm$0.299	&	97.404$\pm$0.368	&	HRS	\\
2011 Sep 09	&	55813.703168	&	55.707$\pm$0.848	&	-	&	APG	\\
2011 Sep 19	&	55823.727209	&	6.791$\pm$0.774	    &	-	&	APG	\\
2011 Oct 15	&	55849.579097	&	63.614$\pm$0.666	&	-	&	APG	\\
2011 Oct 17	&	55851.649485	&	43.595$\pm$0.619	&	-	&	APG	\\
2011 Nov 01	&	55866.570111	&	94.037$\pm$0.683	&	-	&	APG	\\
\hline
\end{tabular}
\end{center}
\end{table}

\subsection{Binary Star Modeling}

For each binary star in our sample we simultaneously modeled the \kep\ lightcurve, and APOGEE and HRS radial velocity curves to determine accurate binary star parameters. To do this we used a combination of software: the binary star modeling software, \ph\ 1.0 \citep{Prsa2005}, which is based on the Wilson-Devinney code (\citealt{Wilson1971}, hearafter WD); \emcee, a \python\ implementation of the affine invariant Markov chain Monte Carlo sampler proposed by \citet{Goodman2010} and implemented by \citet{DFM2013}; \celerite \citep{DFM2017}, the Gaussian process library \citep{DFM2017}; and our own codes. We briefly describe the analysis here, but refer the reader to M19 for a more detailed description.

\subsection{The \ph\ Model}
\label{sec:ph_model}

The \ph\ modelling software combines the complete treatment of the Roche potential with the detailed treatment of surface and horizon effects such as limb darkening, reflection and gravity darkening to derive an accurate model of the binary system. \ph\ uses the WD method of summing over the stellar surface using discrete trapezoidal elements to determine the total observed flux and consequently a complete set of stellar parameters. Additional to the standard functionality of WD, \ph\ offers features including an intuitive graphical user interface, updated filters, and interfacing between \ph\ and \python.

As \ph\ is computationally expensive when used within a Bayesian framework, we elected to compute the \ph\ model in phase space, which significantly reduced the number of data points and thus computational time. This is appropriate as the selected binary systems do not present any temporal variations on timescales longer than that of the orbit (with the exception of spots, which we model using Gaussian Processes in time space). Using \ph, we simultaneously model the \kep\ light curve, and APOGEE and HRS radial velocities. For each system we fit the following parameters: inclination $i$; eccentricity $e$; argument of periastron $\omega$; the primary and secondary surface potentials, $\Omega_1$ and $\Omega_2$; barycentric gamma velocity, $\gamma$; the mass ratio $q$; semi-major axis $a$; third light $l_3$; and the effective temperature ratio $T_{\mathrm{eff}2}/T_{\mathrm{eff}1}$ (where the primary star temperature was fixed and the ratio of the temperatures was used as a fitted parameter due to its orthogonal nature). To ensure that our uncertainty estimates are accurate, we marginalize over the albedos, gravity darkening parameters and rotational velocities, $v\sin i$, for both components.

The selected limb darkening law primarily affects the fit at the ingress and egress portion of the eclipses, thus has implications for the determination of the radii. It is currently unclear which is the preferred limb-darkening law \citep{Maxted2020}, thus we applied the square root limb darkening law to our models, which has been shown by \citet{Diaz-Cordoves1992} to work best for objects that radiate towards the IR.

\subsection{Markov chain Monte Carlo Methods}

We incorporated \emcee, a Markov chain Monte Carlo (McMC) sampler, to sample the parameters' posterior probability distributions within a Bayesian framework. The log-likelihood distribution function was calculated at each iteration:
\begin{equation}
\log{P(\theta|D)} = \log{P(F|\theta)} + \log{P(RV|\theta)} + \log{P(GP|\theta)} + \log{P(\theta)} + C,
\end{equation}
where $D$ denotes the data, $F$ are the light curve measurements, $RV$ are the radial velocity measurements, GP denotes the Gaussian process noise model, $\theta$ is the parameter vector that contains the fitted parameters (specified in \S\ref{sec:ph_model}) and $C$ is an arbitrary constant. A significant benefit of MCMC is that the results are given as posterior probability distributions. By considering these distributions, and subsequently the joint parameter distributions, we are able to see how well the parameter values are determined with respect to the data set.

For each of the fitted parameters in \S\ref{sec:ph_model} we started with an initial distribution which represents a tight $N$-dimensional cube around all parameters in phase space. For each parameter, we applied a uniform prior that encompassed all allowable configurations and additionally marginalized over $v\sin i$ for both components. To constrain our models, we elected to use 288 walkers during the initial burn in time, while the model was visibly converging, and 144 walkers for the remaining duration. We elected to use 288 and 144 walkers as the number of walkers can be equally divided between the available processors (72), which optimizes the time required per iteration (each processor generates 4 or 2 models per iteration, respectively). To estimate convergence, we required a minimum of ten auto-correlation timescales, as recommended by \citet{DFM2013}, although in practice we typically acquired greater than 50.

\subsection{Noise modeling using Gaussian Processes\label{sec:GP}}

We elected to model the systematic noise and unwanted astrophysical signals, such as spots, using the Gaussian Processes (GP) package \celerite. A Gaussian process is a collection of random variables, any finite number of which have joint Gaussian distributions. Within our model the GPs were conditioned on the data so that they were distributed normally with respect to the data. Rather than fitting a function to the noise in our data, we applied GPs to determine the probability distribution over all possible functions that fit the noise. By fitting the high-level parameters of a covariance kernel, GPs enabled a better treatment of noise and poorly understood error correlations in the heteroscedastic data.

Once the model light and radial velocity curves had been created using \ph\ and \emcee, we applied GPs to the model light curve in time space. At the heart of GPs is the kernel or covariance function, which encodes our assumptions about the nearness or similarity between data points. For more information on covariance functions, we direct the interested reader to Chapter 4 of \citet{Rasmussen2006}. For our work we selected a term that is a Taylor-series expansion of the Mat\'{e}rn 3/2 kernel function. The Mat\'{e}rn 3/2 has rapidly and slowly varying components and thus is optimal for modeling long term trends such as spots and more rapid features such as instrumental variations:

\begin{equation}
\label{eqn:matern}
k(\tau) = \sigma^2 \left[ (1+1/\epsilon)e^{-(1-\epsilon)\sqrt{3}\tau/\rho} (1+1/\epsilon)e^{-(1+\epsilon)\sqrt{3}\tau/\rho}\right]
\end{equation}

where $\sigma$ and $\rho$ are the only two tunable parameters and $\epsilon$ controls the quality of the approximation since, in the limit as ${\epsilon \to 0}$, Equation (\ref{eqn:matern}) becomes the Mat\'{e}rn-3/2 function. 

In \S\ref{sec:kepler} we detailed the method used to determine the uncertainties in the light curve. As an additional measure, we applied the jitter kernel to our model to assess the designated uncertainties. The jitter kernel takes the form:

\begin{equation}
\label{eqn:noise}
k(\tau_{n,m}) = \sigma^2\delta_{n,m}, 
\end{equation}

where $\sigma$ is a tunable parameter. In the case that our uncertainties are underestimated, the value of $\sigma$ is significant with respect to the uncertainties and white noise is added to the model. We primarily apply the jitter kernel to assess the uncertainties determined in \S\ref{sec:kepler}.

\subsection{Comparison with Stellar Evolutionary Models}

To verify our results, validate the stellar evolution models and to determine the age of each binary system, we generate isochrone models for each binary star and showed that, for each binary star analysed, both components lie on the same isochrone. We use the {\sc {mist}} ({\sc {mesa}} Isochrones and Stellar Tracks) software \citep{Dotter2016,Choi2016}, which is based on the {\sc {mesa}} (Modules for Experiments in Stellar Astrophysics) package \citep{Paxton2011,Paxton2013,Paxton2015}. The models include time-dependent, diffusive, convective overshoot \citep{Herwig2000}, \citet{asplund2009} solar abundances and the OPAL \citep{Iglesias1993, Iglesias1996} opacities. For each system we used the metallicity and primary star $T_{\rm eff}$ determined from \apogee\ data release 16 \citep{Jonsson2020}.  We are able to assume that the \apogee\ temperatures are equivalent to the primary temperatures as all systems have a small H-band flux ratio, which infers that the contribution of the secondary component to the determined temperature is not substantial.

\section{Fundamental Parameters for four benchmark-grade Eclipsing Binaries \label{sec:benchmark}}

For each binary star, the precise radial velocities and \kep\ light curves have been modeled using \ph, \emcee\ and \celerite. The results comprise accurate masses and radii for all components, shown in Table\,\ref{tab:results}. 
\begin{table}{
\caption{
\label{tab:results}
Model parameters for the four benchmark eclipsing binaries}	
\begin{center}
\begin{tabular}{l r r r r}
\hline
Parameter	&	KIC\,2306740	&	KIC\,4076952	&	KIC\,5193386	&	KIC\,5288543	\\
\hline										
\multicolumn{5}{l}{\emph{DR16 \apogee\ parameters}} \\
\hline
$T_\mathrm{A}$ (K)	&	5570$\pm$140	&	6590$\pm$220	&	4770$\pm$100	& 6280$\pm$170 \\
$[$Fe/H$]$ &	-0.59$\pm$0.02	&	-0.06$\pm$0.04	&	-0.50$\pm$0.01	&	-0.55$\pm$0.03\\
\hline
\multicolumn{5}{l}{\emph{Derived Orbital Parameters}}\\		\hline
Period	&	10.30699$\pm$0.00003	&	9.76116$\pm$0.00002	&	21.37829$\pm$0.00009	&	3.457075$\pm$0.000006	\\	
T$_0$	&	54966.425$\pm$0.067	&	54966.703$\pm$0.073	&	54980.20$\pm$0.17	&	54964.805$\pm$0.032	\\	
$i$ (deg)  	&	88.809$\pm$0.003	&	88.915$\pm$0.002	&	88.921$\pm$0.002	&	88.183$\pm$0.0002	\\	
$e$	&	0.3072$\pm$0.004	&	0.0306$\pm$0.0001	&	0.0092$\pm$0.002	&	0.0030$\pm$0.0001	\\	
$\omega$ (rad)	&	4.7881$\pm$0.0001	&	0.958$\pm$0.003	&	0.8016$\pm$0.0015	&	3.86$\pm$0.04	\\	
$\Omega_A$	&	16.56$\pm$0.01	&	10.007$\pm$0.007	&	10.67$\pm$0.01	&	8.52$\pm$0.01	\\	
$\Omega_B$	&	20.69$\pm$0.14	&	15.64$\pm$0.04	&	29.0$\pm$0.4	&	11.34$\pm$0.09	\\	
$l_3$ (\%)	&	0.02$\pm$0.025	&	0.033$\pm$0.001	&	0.007$\pm$0.004	& 0.002$\pm$0.002	\\	
$K_A$	&	63.76$\pm$0.29	&	62.6$\pm$0.3	&	47.49$\pm$0.33	&	78.85$\pm$0.43	\\	
$K_B$	&	72.14$\pm$0.44	&	84.8$\pm$0.5	&	57.1$\pm$0.7	&	104.14$\pm$0.61	\\	
$\gamma$	&	15.10$\pm$0.18	&	18.1$\pm$0.3	&	-40.04$\pm$0.23	&	29.6$\pm$0.3	\\	
$q$	&	0.884$\pm$0.006	&	0.738$\pm$0.002	&	0.831$\pm$0.011	&	0.757$\pm$0.007	\\	
$a$ ($\Rsun$)	&	26.35$\pm$0.11	&	28.42$\pm$0.14	&	44.2$\pm$0.3	&	12.51$\pm$0.04	\\	
\hline
\multicolumn{5}{l}{\emph{Physical Parameters}} \\			\hline
$M_A$ ($\Msun$) 	&	1.227$\pm$0.017	&	1.86$\pm$0.03	&	1.386$\pm$0.036	&	1.251$\pm$0.015	\\	
$M_B$ ($\Msun$) 	&	1.085$\pm$0.012	&	1.37$\pm$0.02	&	1.152$\pm$0.020	&	0.947$\pm$0.010	\\	
$R_A$ ($\Rsun$)	&	1.734$\pm$0.007	&	3.078$\pm$0.016	&	4.505$\pm$0.032	&	1.6208$\pm$0.0055	\\	
$R_B$ ($\Rsun$)	&	1.215$\pm$0.005	&	1.451$\pm$0.008	&	1.319$\pm$0.009	&	0.9272$\pm$0.0032	\\	
$T_B$/$T_A$	&	0.9655$\pm$0.0004	&	1.0137$\pm$0.0003	&	1.3862$\pm$0.0011	&	0.8905$\pm$0.0001	\\	
$L_A/(L_A+L_B)$	&	0.70224$\pm$0.00033	&	0.81122$\pm$0.00022	&	0.7587$\pm$0.0005	&	0.83010$\pm$0.00008	\\	
\hline
\multicolumn{5}{l}{\emph{Gaussian Process Parameters}} \\	\hline
$\log$($\phi_\mathrm{matern}$)	&	-1.493$\pm$0.011	&	-1.869$\pm$0.013	&	-1.67$\pm$0.02	&	-1.7$\pm$0.2	\\	
$\log$($\rho_\mathrm{matern}$)	&	2.005$\pm$0.005	&	2.89$\pm$0.01	&	2.012$\pm$0.009	&	2.9$\pm$0.19	\\	
$\log$($\phi_\mathrm{jitter}$)	&	-7.485$\pm$0.001	&	-8.973$\pm$0.005	&	-8.9397$\pm$0.0009	&	-7.899$\pm$0.003	\\	
\hline										
\end{tabular} 
\end{center} 	
}
\end{table}

\subsection{KIC\,2306740}

\begin{figure}
\begin{center}
\includegraphics*[width=5.2in,angle=0]{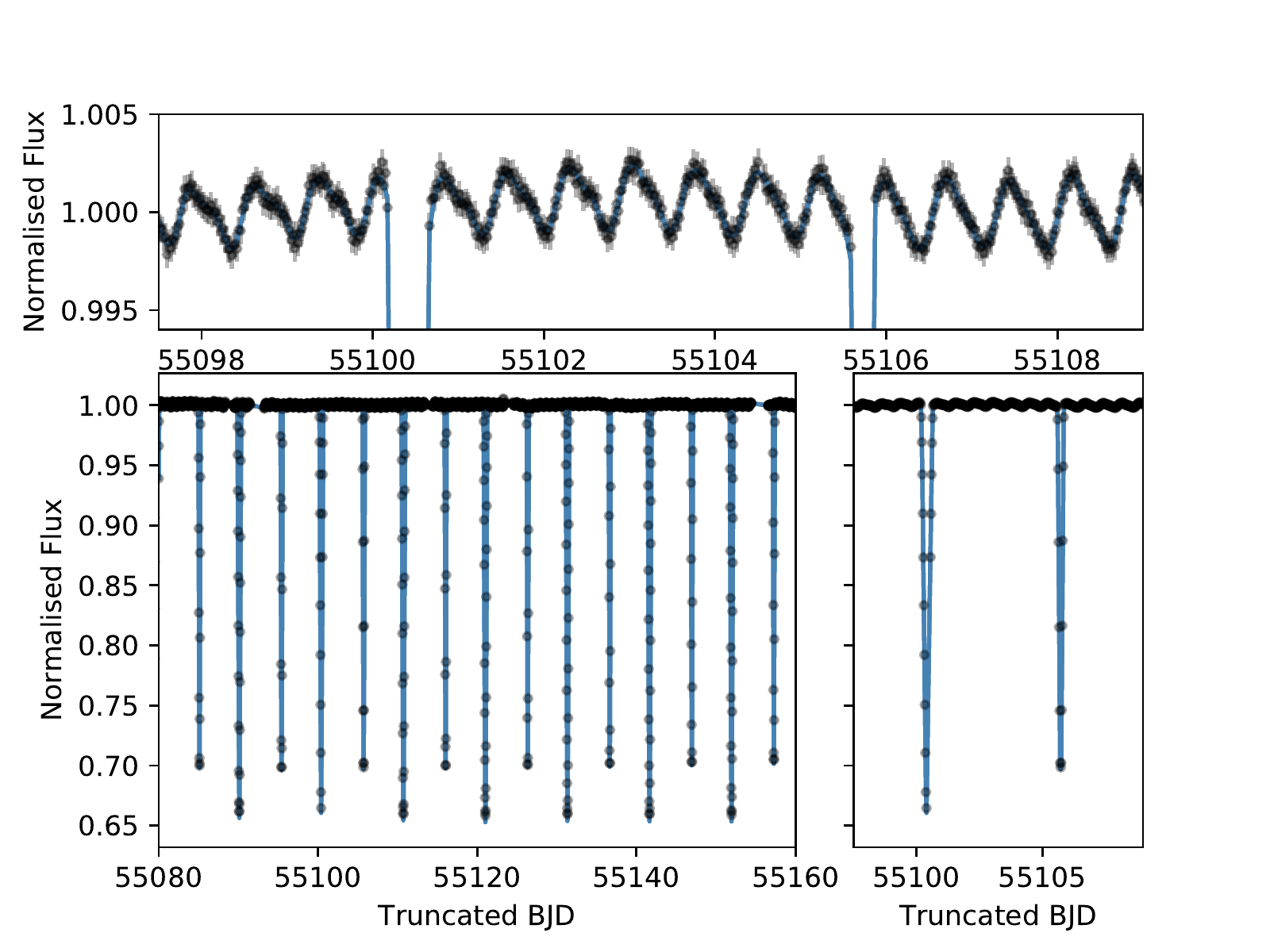}
\end{center}
\caption{The \kep\ light curve of KIC 2306740. Here the black points denote the \kep\ data and the blue lines denote the \ph\ binary star model and Gaussian processes combined. The out of eclipse variation has been attributed to spots.} \label{fig:2306lc}
\end{figure}

KIC\,2306740 consists of two F-type stars in a binary system with a $10.30699 \pm 0.00003$\,d orbital period. The system exhibits a periodic oscillation in the out of eclipse region, as seen in Figure\,\ref{fig:2306lc}. The amplitude and frequency of the oscillation are variable over long timescales (visible on the timescale of the dataset), making pulsations unlikely as the cause of the variation. The Fourier transform exhibits two significant peaks, the second (2.7280\,\cd) the harmonic of the first (1.36387\,\cd), indicative of spots \citep{Lanza1994,Zhan2019}. If that is indeed the case, one object is rotating 3.8 times per orbit, consistent with our findings for the secondary object (from the binary model) to be rotating at a rate of $2.8\pm1.3$ times per orbit.

\begin{figure}
\begin{center}
\includegraphics*[width=5.2in,angle=0]{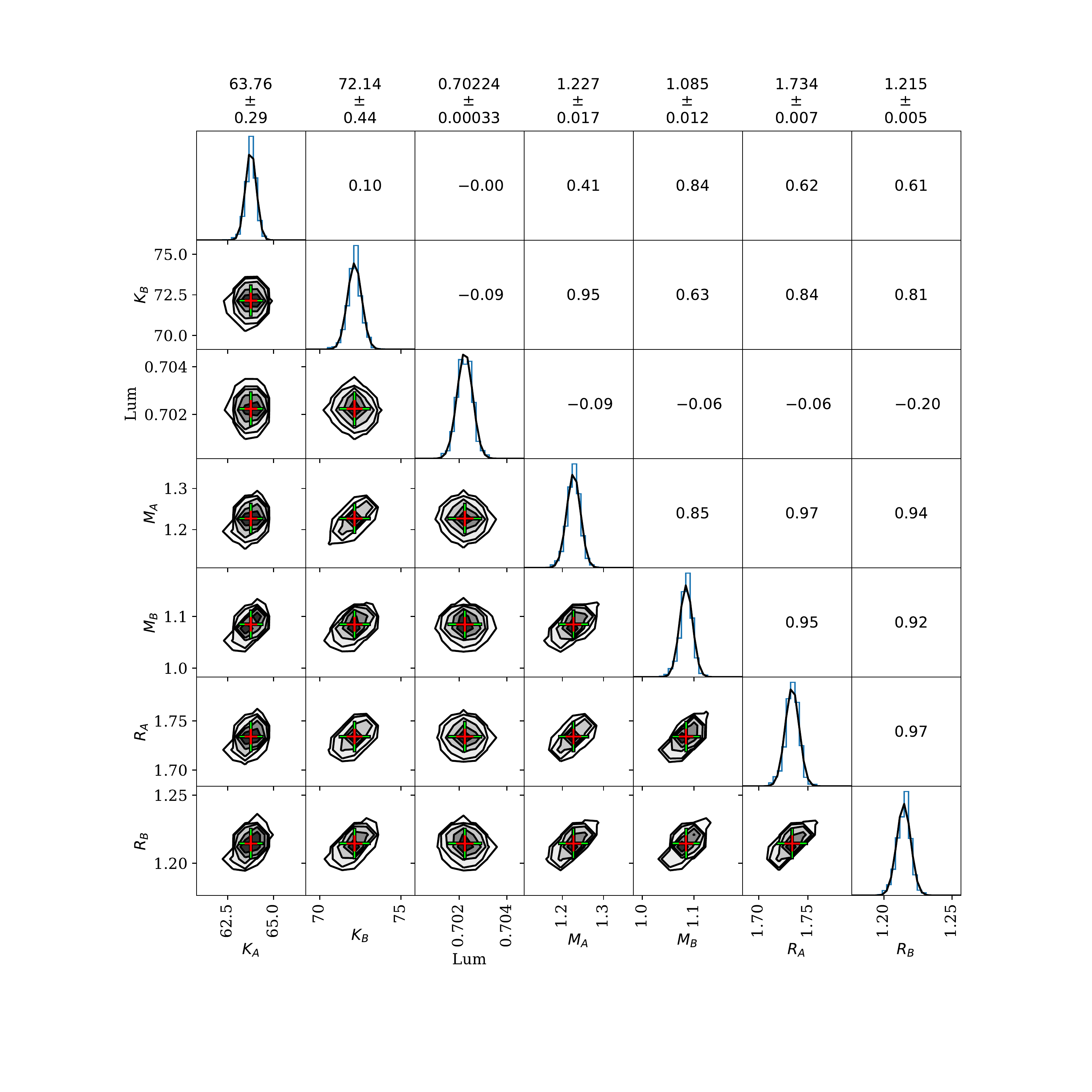}
\end{center}
\caption{Distributions for the calculated parameters of KIC\,2306740 where K is the semi-amplitude, Lum is the Luminosity ratio ($L_A/(L_A+L_B$), and M and R are the Mass and Radius, respectively. The lower left section shows the 2-D parameter cross sections and the values reflected about the diagonal are the correlation parameter associated with each cross section. The histograms on the diagonal represent the distributions for each individual parameter and the values across the top are the mean and standard deviation based on the Gaussian fit to the histograms.\label{fig:2306blobs}} 
\end{figure}

\begin{figure}
\begin{center}
\includegraphics*[width=5.2in,angle=0]{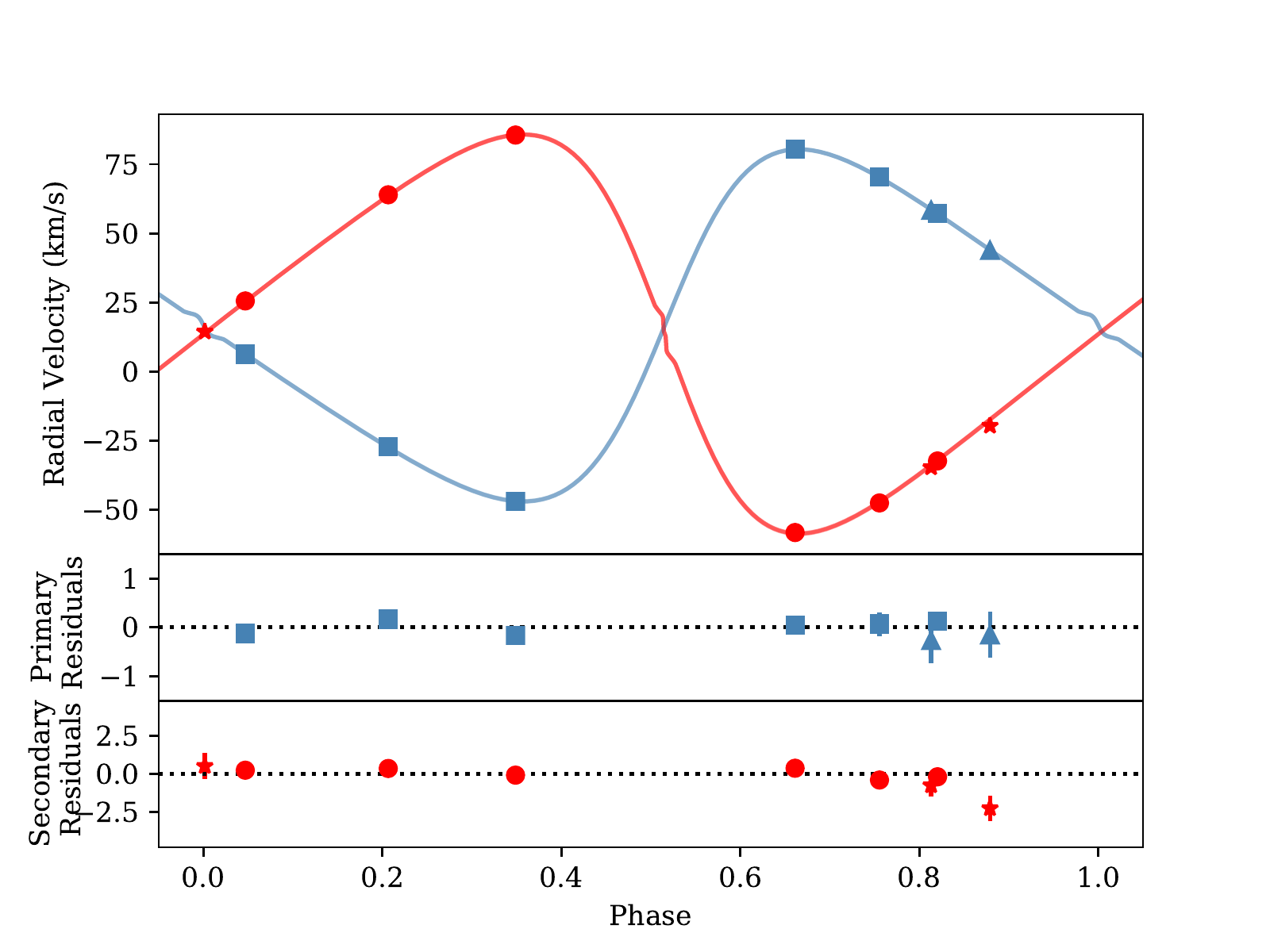}
\end{center}
\caption{The figure shows the APOGEE and HRS radial velocities for KIC\,2306740. Here the HRS radial velocities are denoted by blue squares (primary) and red circles (secondary), and APOGEE observations are denoted by the blue triangles (primary) and red stars (secondary). The blue and red lines denote the \ph\ model for the primary and secondary components, respectively. The lower panels depict the residuals for the primary and secondary components. Note the Rossiter-McLaughlin effect is visible for both stars as they are eclipsed.\label{fig:2306rvs}}
\end{figure}

The fundamental parameters of KIC 2306740 were determined to be: $M_A = 1.227\pm 0.017\,\Msun$, $M_B = 1.085\pm 0.012\,\Msun$, $R_A = 1.734\pm 0.007\,\Rsun$ and $R_B = 1.215\pm 0.005\,\Rsun$. The distributions of these parameters can be seen in Figure\,\ref{fig:2306blobs}. The exceptionally low ($<1.5\%$) uncertainties can be attributed to the combination of the precise \kep\ light curve (see Figure\,\ref{fig:2306lc}) and the precise radial velocity measurements that sample the orbital phase very well (see Figure\,\ref{fig:2306rvs}).

\begin{figure}
\begin{center}
\includegraphics*[width=5.2in,angle=0]{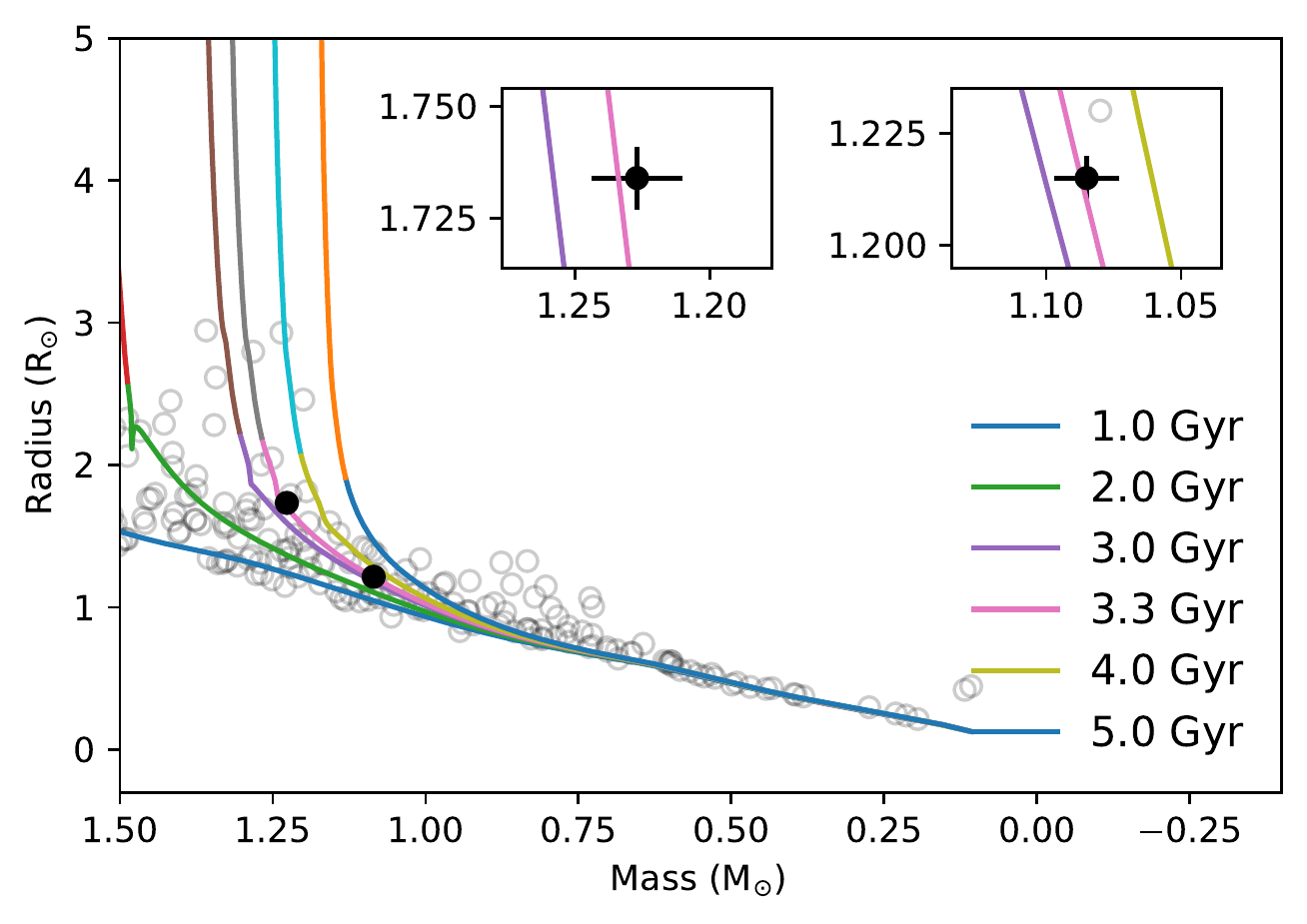}
\end{center}
\caption{Radius versus mass, plotted for the EBs in the DEBCat
  Eclipsing Binary Catalog (open circles), an updated version of the \citet{Andersen1991} catalog, along with the two components of KIC\,2306740 (solid black circles).  For comparison, we show {\sc{mist}} ({\sc{mesa}} Isochrones and Stellar Tracks) stellar isochrones (\citealt{Dotter2016,Choi2016}). The inserts depict magnified regions for the primary and secondary components. \label{fig:2306iso}}
\end{figure}

Even with the small uncertainties, the two components of KIC\,2306740 land on the same isochrone (see Figure\,\ref{fig:2306iso}), which places the two stars at $\sim$3.3\,Gyr with a metallicity of [Fe/H]$ = -0.59$, as determined by the DR16 catalog ($-0.59\pm 0.02$).

\subsection{KIC\,4076952}

\begin{figure}
\begin{center}
\includegraphics*[width=5.2in,angle=0]{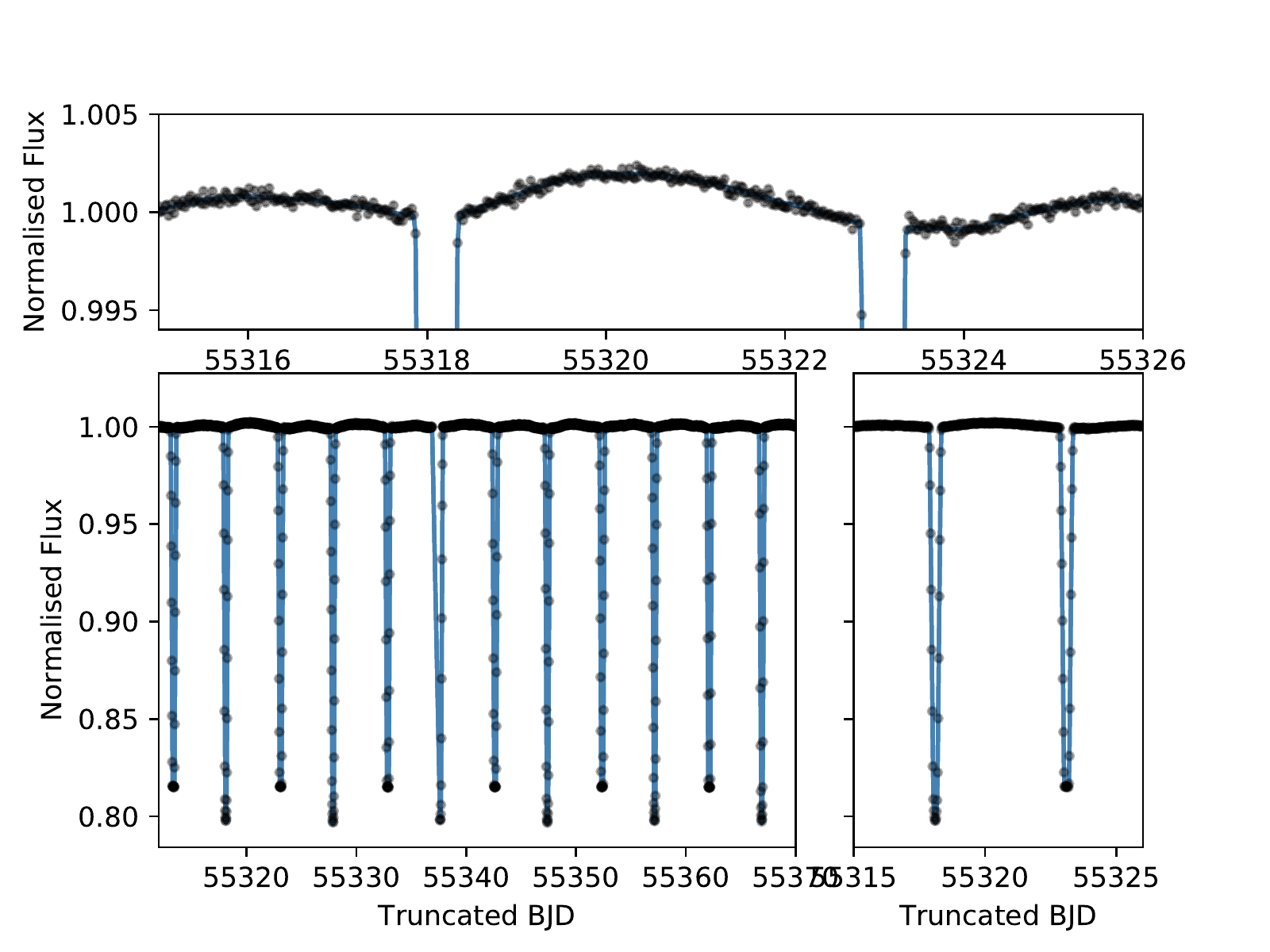}
\end{center}
\caption{The \kep\ light curve of KIC\,4076952. Here the \kep\ data are displayed as black points and the blue lines denote the \ph\ binary star model and Gaussian processes combined. \label{fig:4076lc}}
\end{figure}

\begin{figure}
\begin{center}
\includegraphics*[width=5.2in,angle=0]{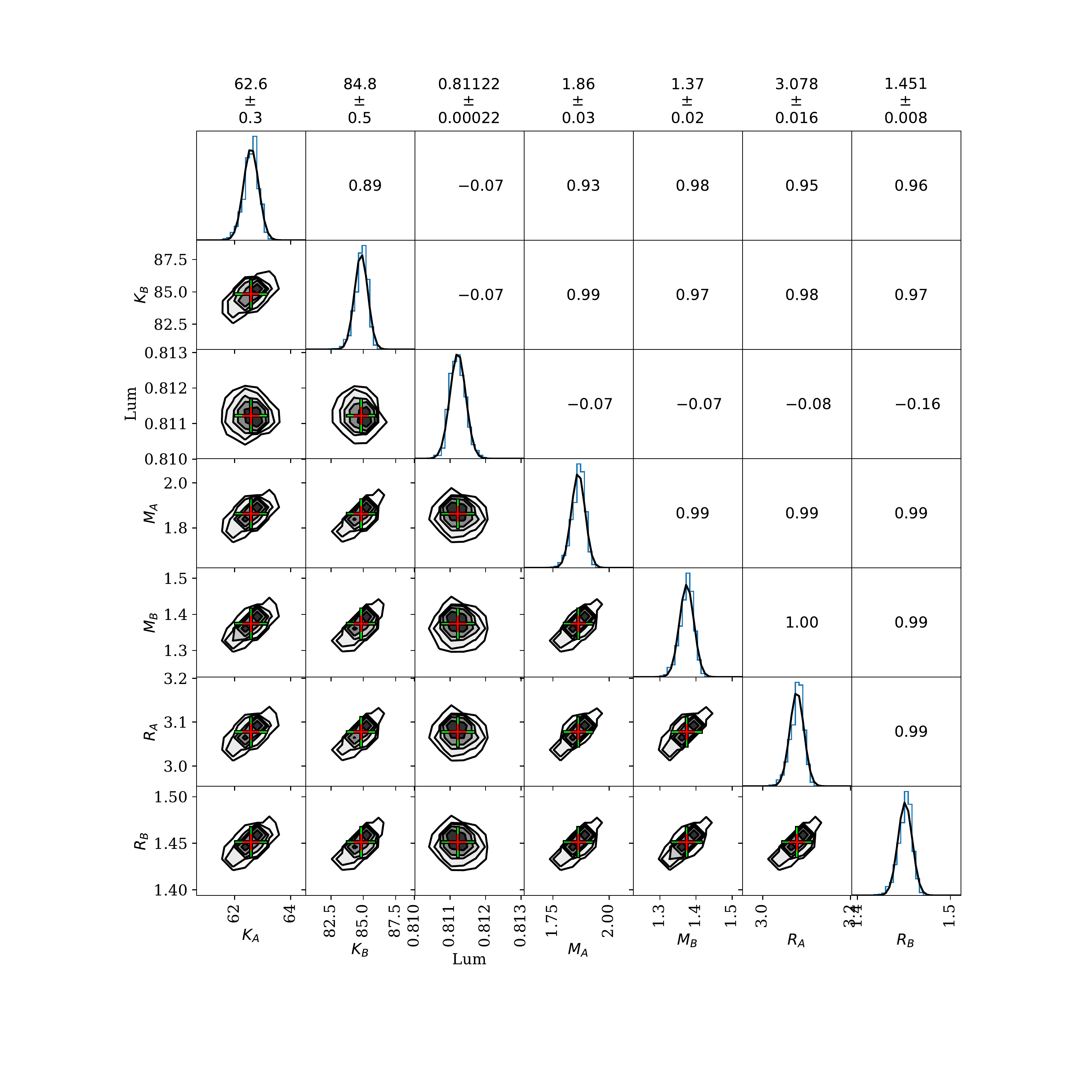}
\end{center}
\caption{Distributions for KIC\,4076952 where the figure has the same format as Figure\,\ref{fig:2306blobs} \label{fig:4076blobs}}
\end{figure}

\begin{figure}
\begin{center}
\includegraphics*[width=5.2in,angle=0]{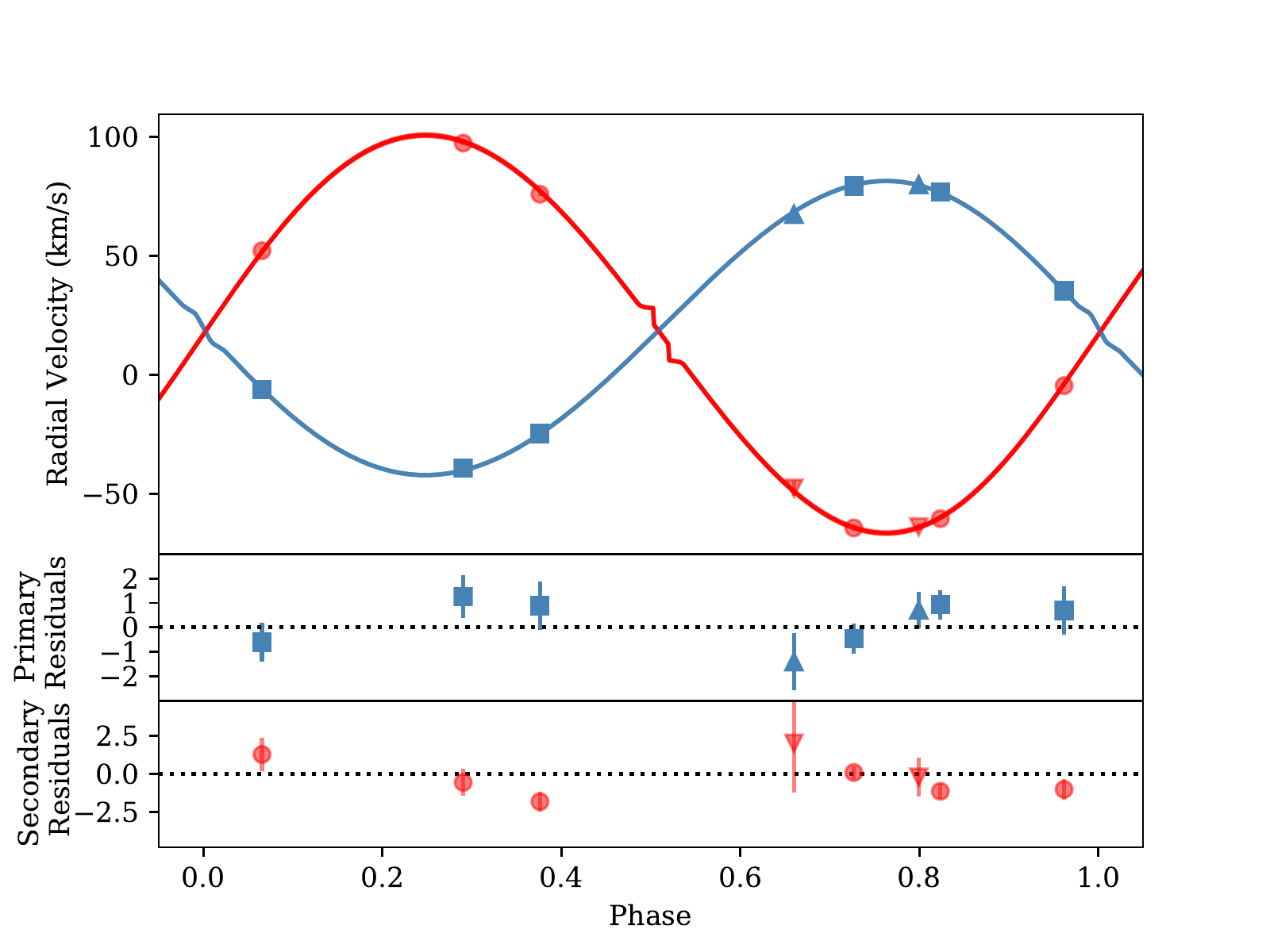}
\end{center}
\caption{The figure depicts the \apogee\ and HRS radial velocities, and \ph\ model radial velocity curves for KIC\,4076952. The residuals show significant scatter with respect to the derived \todcor\ uncertainties, we attribute this to the uncertainties being slightly underestimated. The format is the same as that of Figure\,\ref{fig:2306rvs}. \label{fig:4076rvs}}
\end{figure}

KIC\,4076952 contains a mid A- and early F-type star in a $9.76116\pm0.00002$\,d orbit. The out-of-eclipse variations seen in the light curve (Figure\,\ref{fig:4076lc}) can be attributed to ellipsoidal variations, which are appropriately fitted within \ph's Roche framework. The fundamental parameters, provided as distributions in Figure\,\ref{fig:4076blobs}, were determined to be: ${{M_A = 1.86\pm 0.03\,\Msun}}$, ${{M_B = 1.37\pm 0.02\,\Msun}}$, ${{R_A = 3.078\pm 0.016\,\Rsun}}$ and ${{R_B = 1.451\pm 0.008\,\Rsun}}$. The parameters have been determined through a simultaneous fit to the light and radial velocity curves (see Figures\,\ref{fig:4076lc} and \ref{fig:4076rvs}, respectively). 

\begin{figure}
\begin{center}
\includegraphics*[width=5.2in,angle=0]{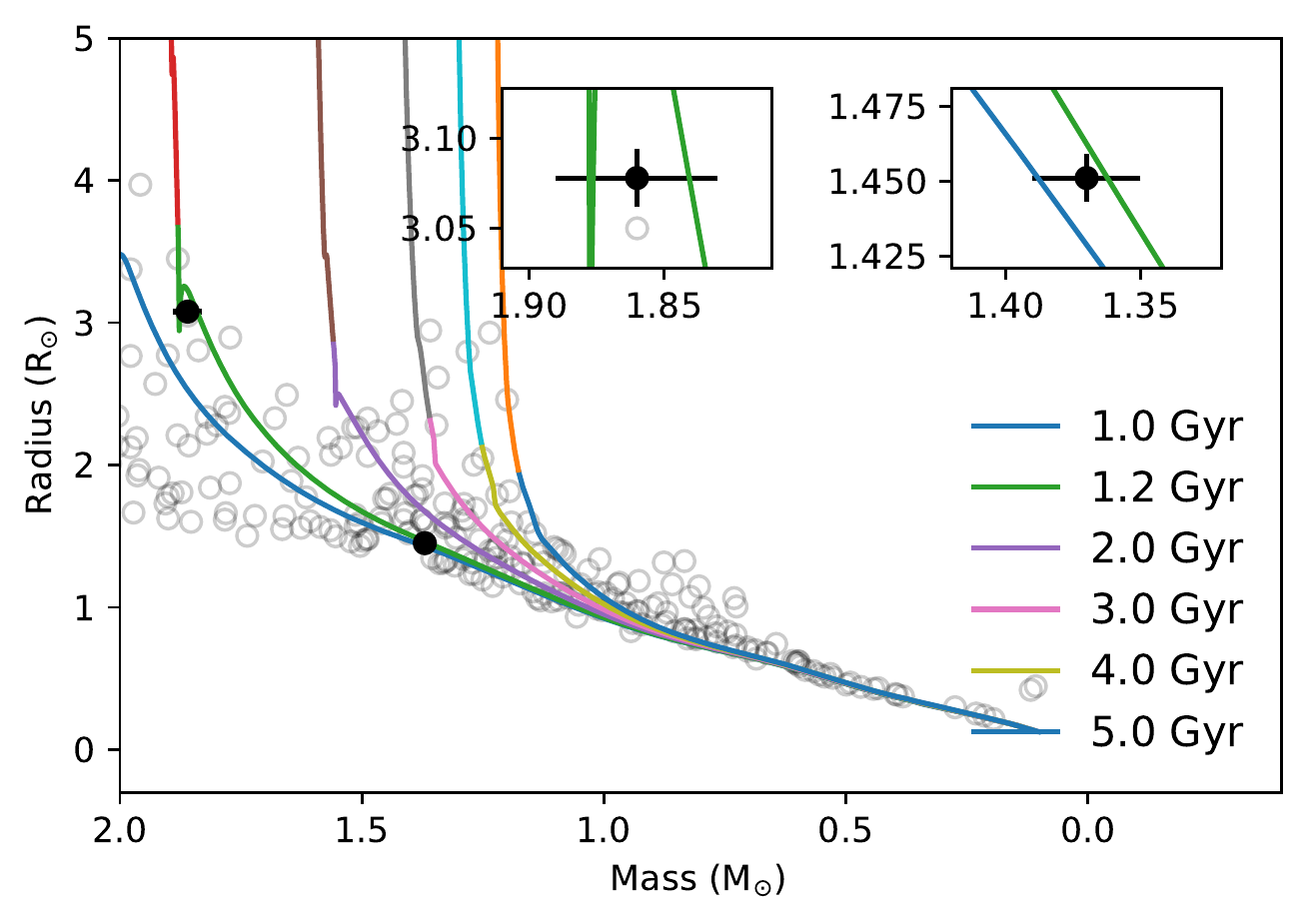}
\end{center}
\caption{Radius vs mass for the components of KIC\,4076952 where the figure is the same as Figure\,\ref{fig:2306iso}. \label{fig:4076iso}}
\end{figure}

The stellar isochrones, which were ascribed a metallicity of [Fe/H]\,=\,-0.06, suggest that the two stars are $\sim$1.2\,Gyr old and that the primary star is starting to evolve off the main sequence (shown by its proximity to the red line in Figure\,\ref{fig:4076iso}). The metallicity determined by the DR16 pipeline is [Fe/H]$ = -0.06 \pm 0.04$.

\subsection{KIC\,5193386}

\begin{figure}
\begin{center}
\includegraphics*[width=5.2in,angle=0]{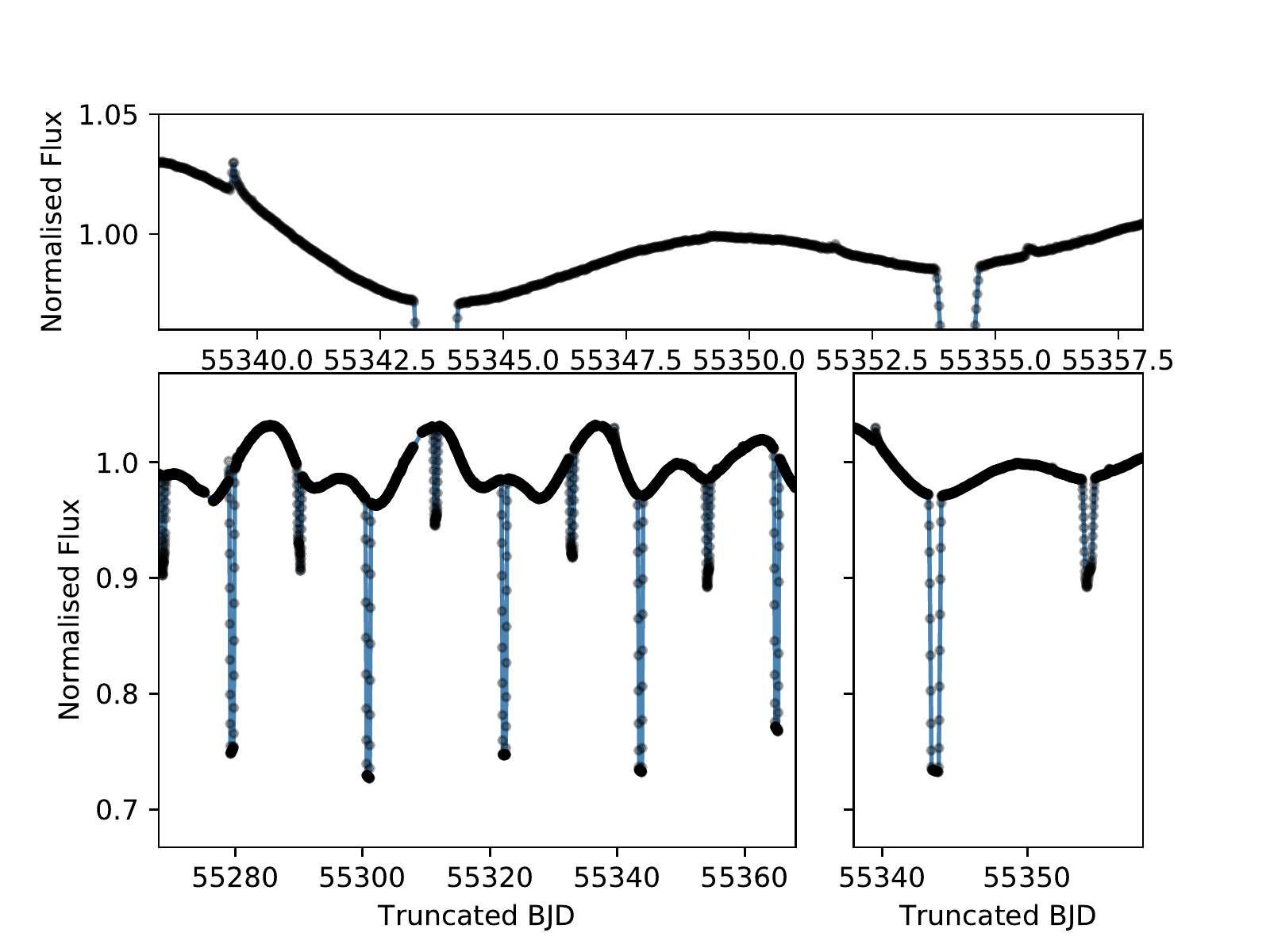}
\end{center}
\caption{The \kep\ light curve of KIC\,5193896. Here the \kep\ data are displayed as black points and the blue lines denote the \ph\ binary star model and Gaussian processes combined. The significant variation in the envelope of the light curve are due to spots on the primary red giant component.} \label{fig:5193lc}
\end{figure}

KIC\,5193386 is an eclipsing binary star in a $21.37829\pm0.00009$\,d orbit with one evolved F-type star and one main sequence F-type star. The obvious spots in the light curve (as seen in Figure\,\ref{fig:5193lc}) are commonly found on red giant stars and thus are attributed to the evolved component. Within our model, spots are treated as astrophysical noise and thus we account for them using Gaussian processes.

\begin{figure}
\begin{center}
\includegraphics*[width=5.2in,angle=0]{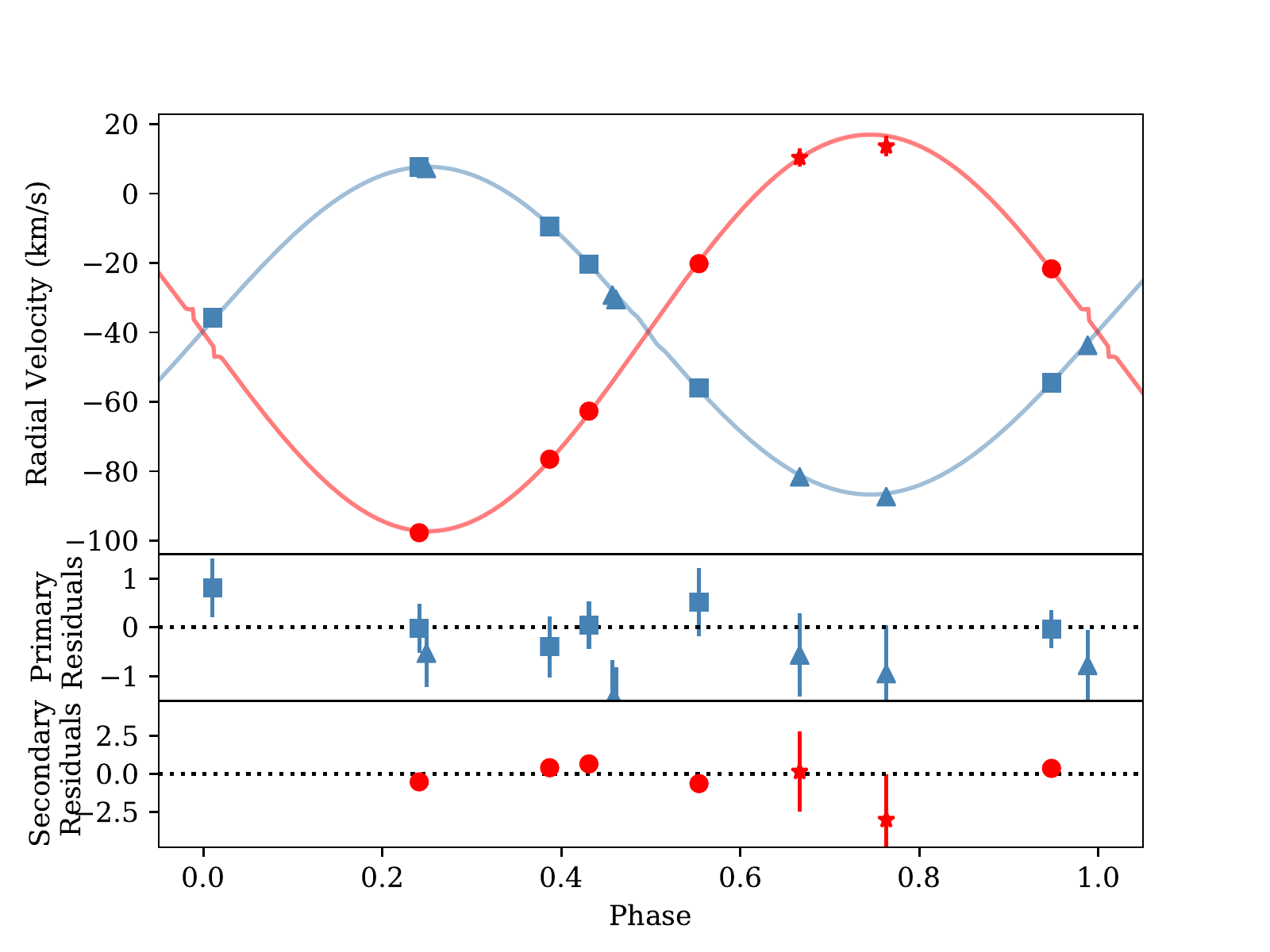}
\end{center}
\caption{The figure depicts the \apogee\ and HRS radial velocities, and \ph\ model radial velocity curves for KIC\,5193386. The format is the same as that of Figure\,\ref{fig:2306rvs}. \label{fig:5193rvs}}
\end{figure}

\begin{figure}
\begin{center}
\includegraphics*[width=5.2in,angle=0]{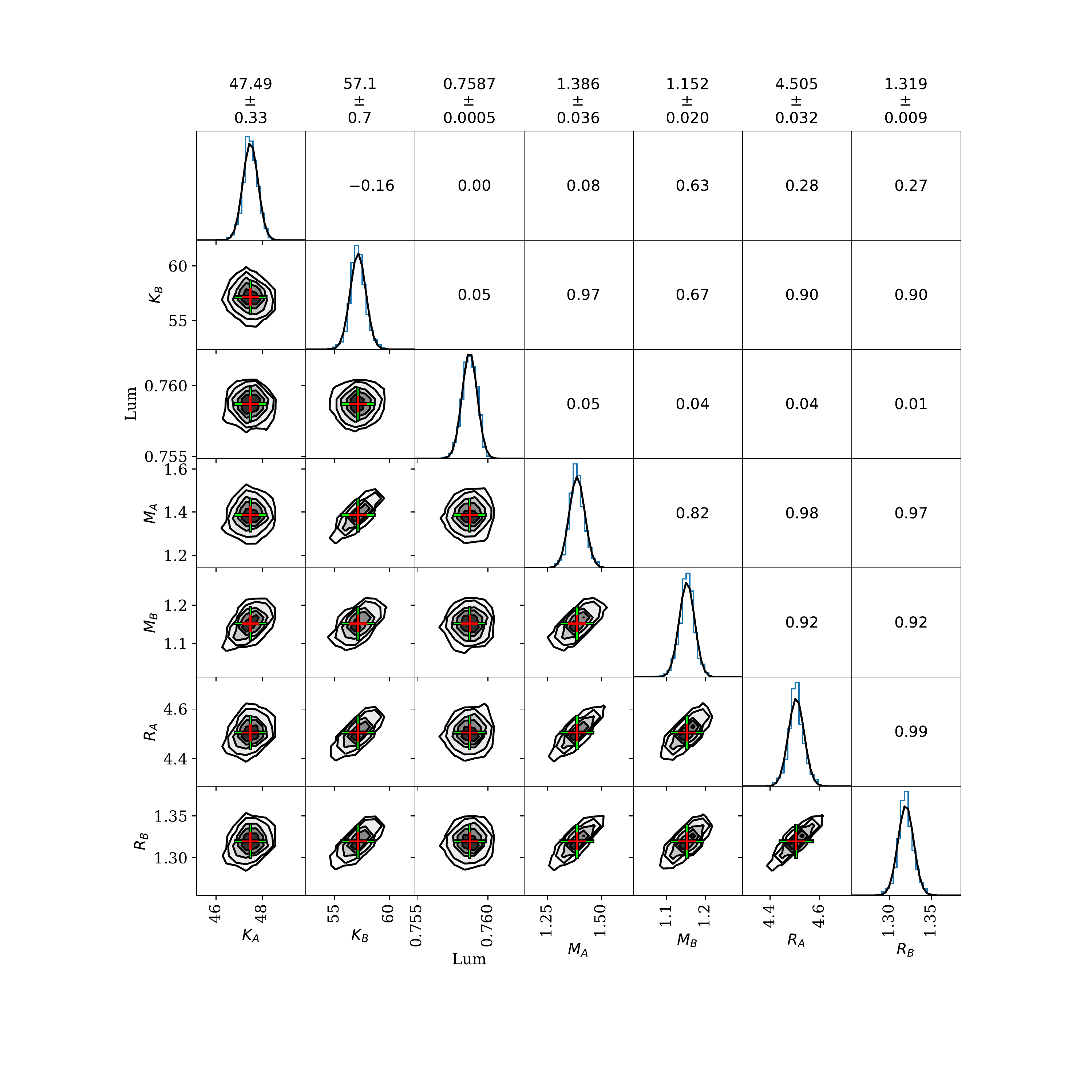}
\end{center}
\caption{Distributions for KIC\,5193386 where the figure has the same format as Figure\,\ref{fig:2306blobs}. \label{fig:5193blobs}}
\end{figure}

From modeling the binary star light and radial velocity curves (as seen in Figures\,\ref{fig:5193lc} and \ref{fig:5193rvs}, respectively), we determined the fundamental parameters to be: $M_A = 1.386\pm 0.036\,\Msun$, $M_B = 1.152\pm 0.020\,\Msun$, $R_A = 4.505\pm 0.032\,\Rsun$ and $R_B = 1.319\pm 0.009\,\Rsun$, which are shown as distributions in Figure\,\ref{fig:5193blobs}.

\begin{figure}
\begin{center}
\includegraphics*[width=5.2in,angle=0]{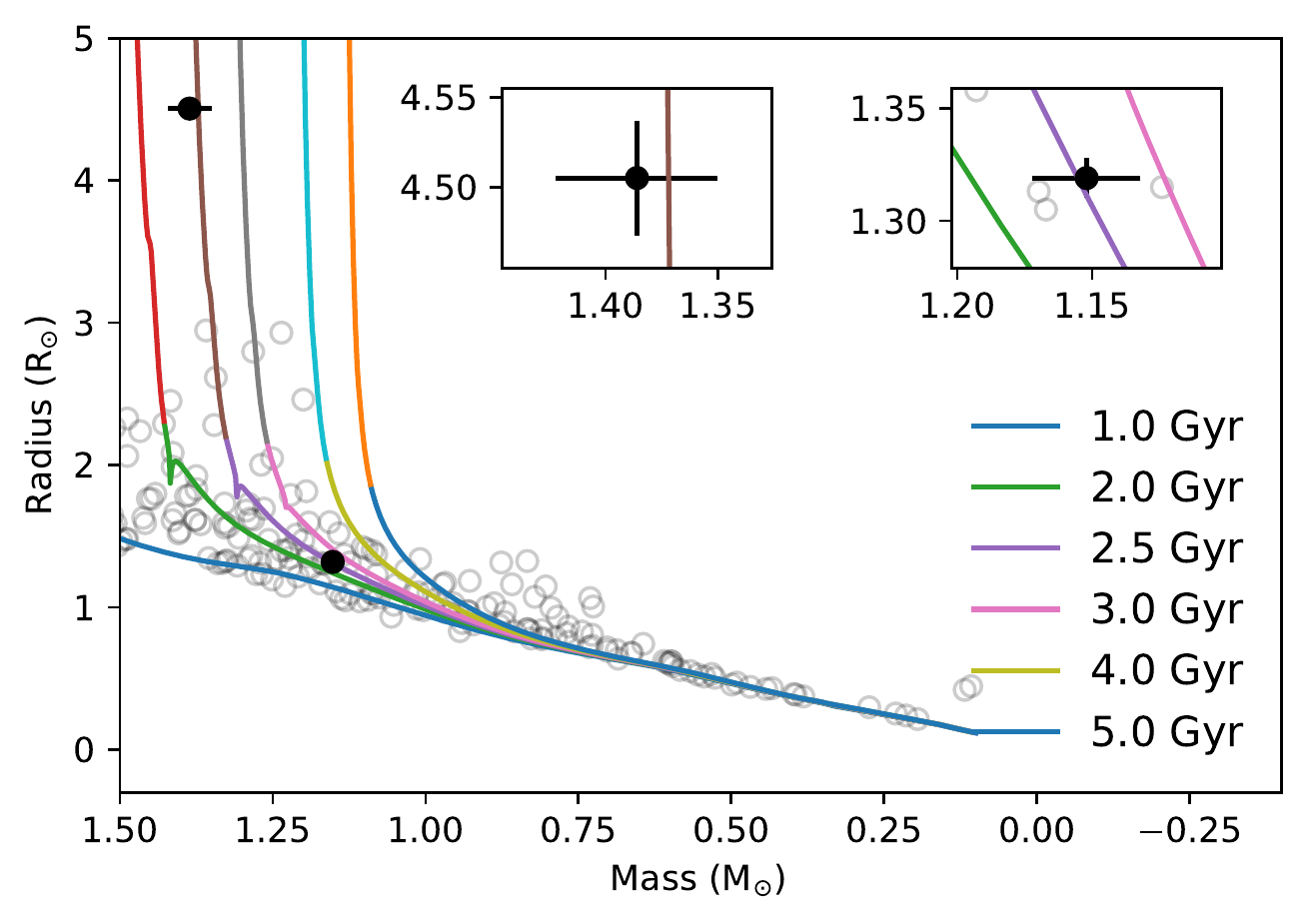}
\end{center}
\caption{Radius vs mass for the components of KIC\,5193386 where the figure is the same as Figure\,\ref{fig:2306iso}.  \label{fig:5193iso}}
\end{figure}

The stellar isochrones, as depicted in Figure\,\ref{fig:5193iso}, show lines of equal age for stars with a metallicity of [Fe/H]\,=\,-0.5 as determined through spectral template fitting, [Fe/H] $= -0.5 \pm 0.01$. Both stars lie on the 2.5\,Gyr isochrone with the more massive star lying on the brown line, which is the red giant branch segment of the isochrone.

\subsection{KIC\,5288543}

\begin{figure}
\begin{center}
\includegraphics*[width=5.2in,angle=0]{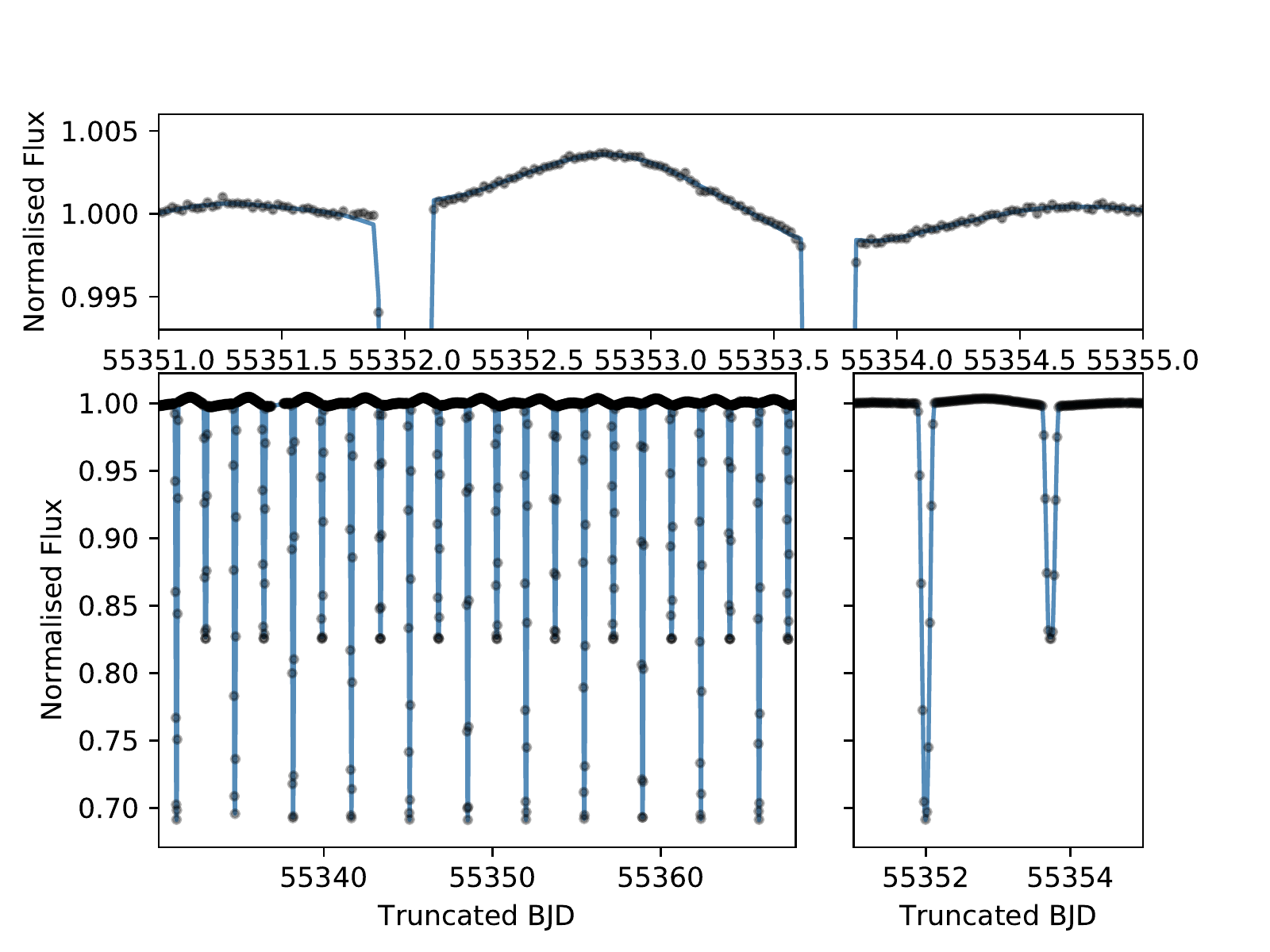}
\end{center}
\caption{The \kep\ light curve of KIC\,5288543. Here the \kep\ data are displayed as black points and the blue lines denote the \ph\ binary star model and Gaussian processes combined. The out of eclipse variations are a combination of spots and ellipsoidal variations.} \label{fig:5288lc}
\end{figure}

KIC\,5288543 consists of two main sequence stars, an F-type and G-type star, in a $3.457075\pm0.000006$\,d binary orbit. The out of eclipse variations in the light curve (see Figure\,\ref{fig:5288lc}) can be attributed a combination of ellipsoidal variations, which are fitted by \ph,\ and spots, which are fitted by gaussian processes. The spots and ellipsoidal variations are easily separated as the spot signal varies significantly over the data set whereas the ellipsoidal variable signal does not.

\begin{figure}
\begin{center}
\includegraphics*[width=5.2in,angle=0]{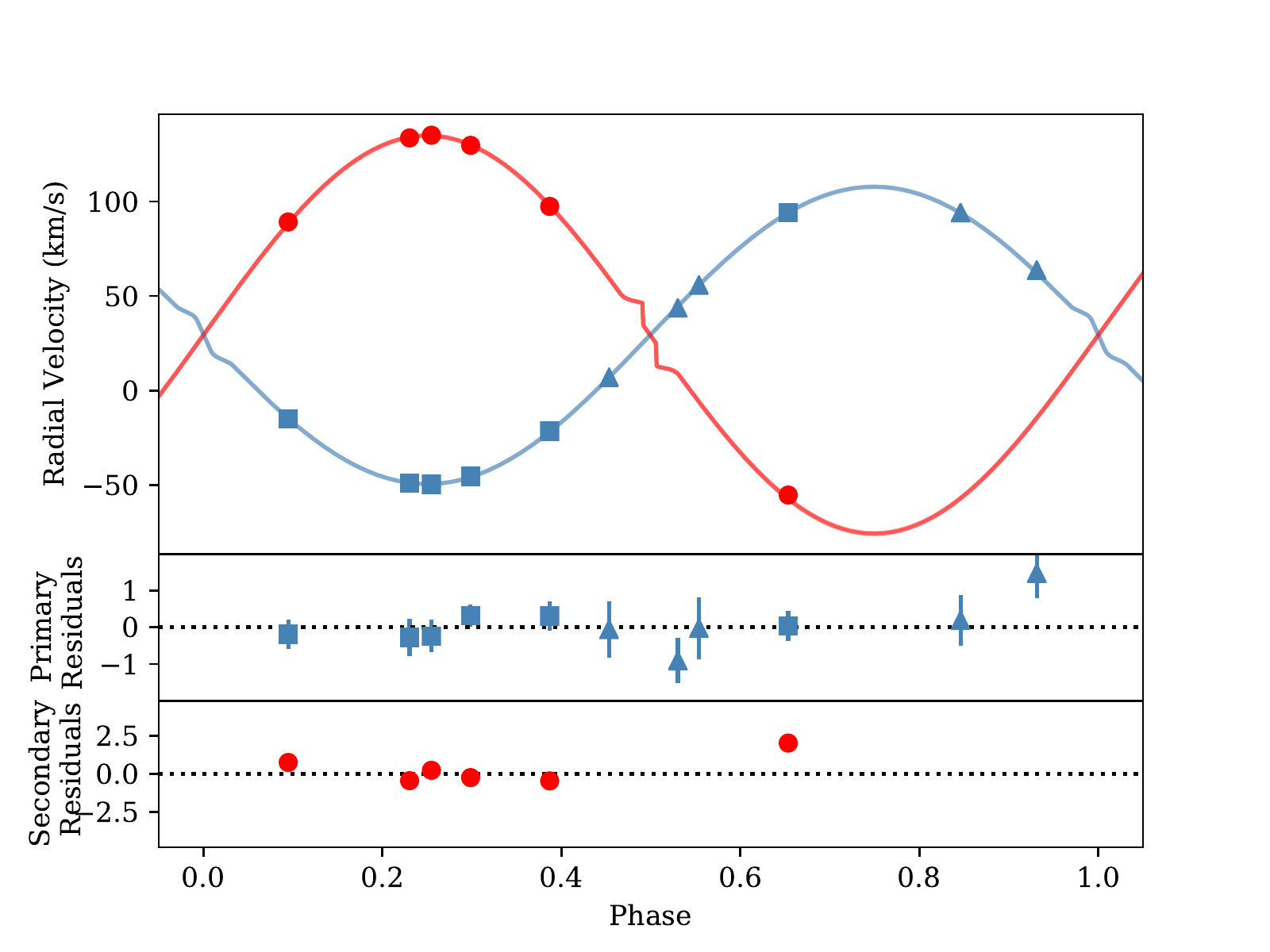}
\end{center}
\caption{The figure depicts the \apogee\ and HRS radial velocities, and \ph\ model radial velocity curves for KIC\,5288543. The format is the same as that of Figure\,\ref{fig:2306rvs}. \label{fig:5288rvs}}
\end{figure}

\begin{figure}
\begin{center}
\includegraphics*[width=5.2in,angle=0]{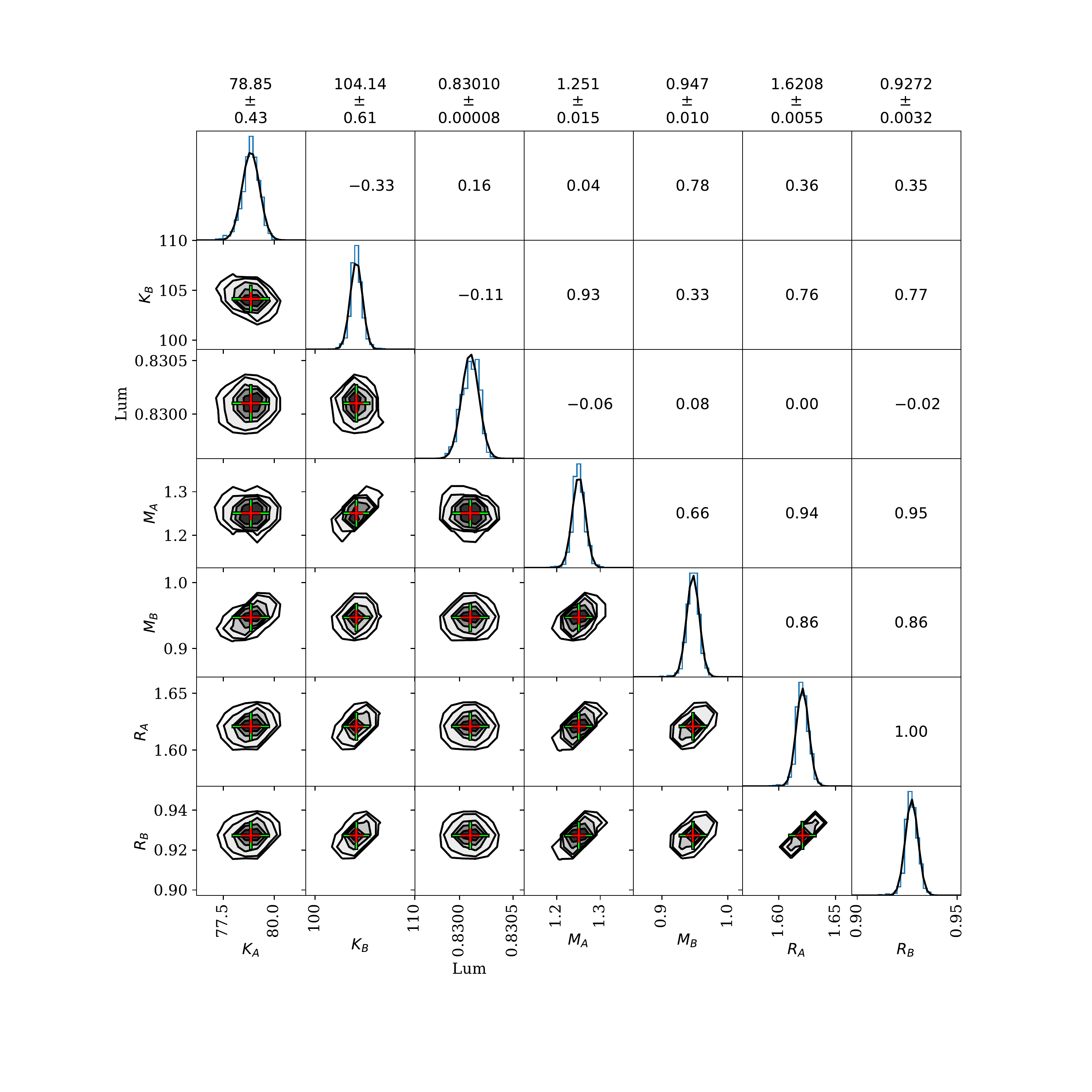}
\end{center}
\caption{Distributions for KIC\,5288543 where the figure has the same format as Figure\,\ref{fig:2306blobs}. \label{fig:5288blobs}}
\end{figure}

The detailed binary light curve and radial velocity curves, as shown in Figures\,\ref{fig:5288lc} and \ref{fig:5288rvs} are well sampled and consequently provide accurate fundamental stellar parameters: $M_A = 1.251\pm 0.015\,\Msun$, $M_B = 0.947\pm 0.010\,\Msun$, $R_A = 1.6208\pm 0.0055\,\Rsun$ and $R_B = 0.9272\pm 0.0032\,\Rsun$, which are depicted as distributions in Figure\,\ref{fig:5288blobs}.

\begin{figure}
\begin{center}
\includegraphics*[width=5.2in,angle=0]{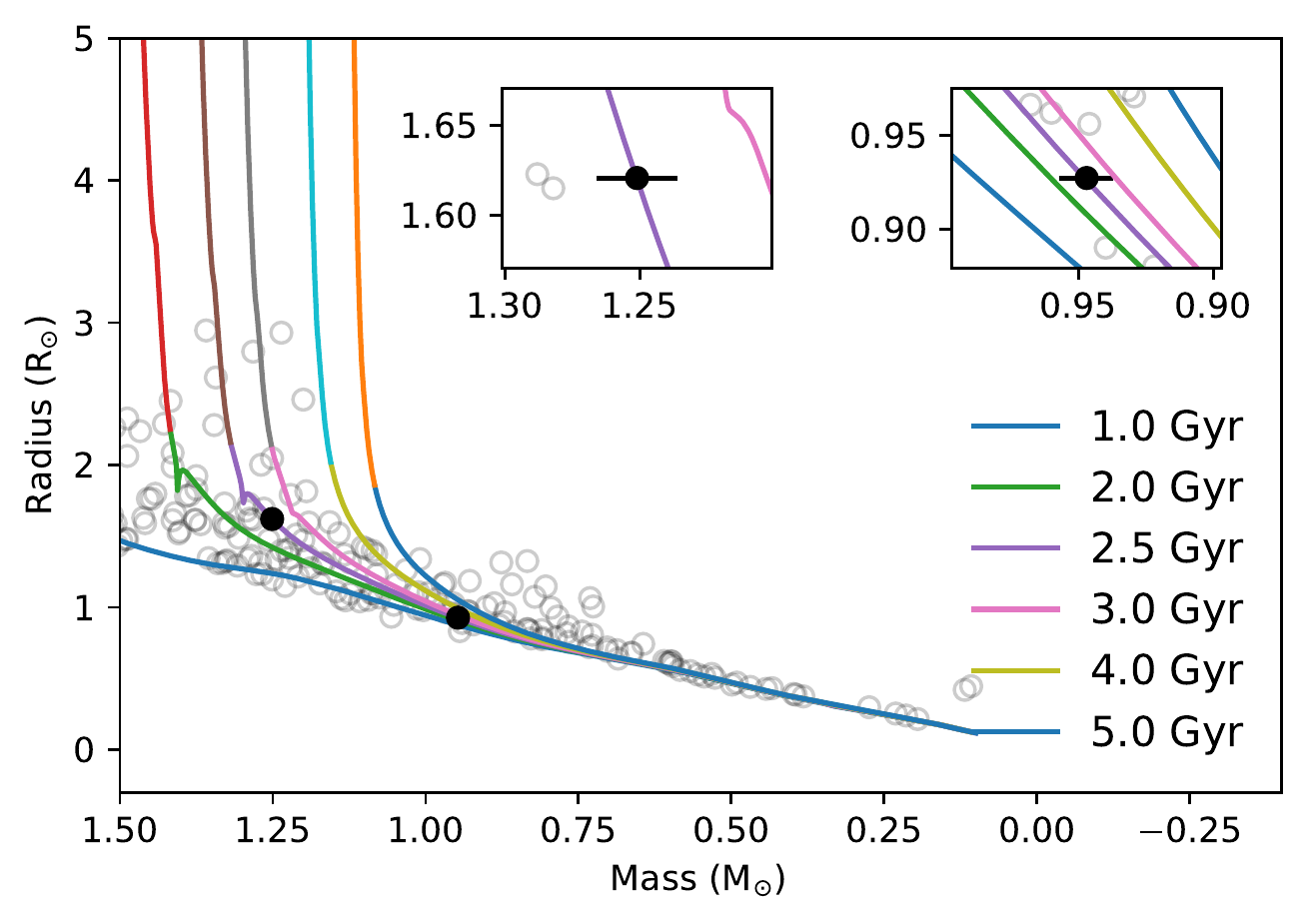}
\end{center}
\caption{Radius vs mass for the components of KIC\,5288543 where the figure is the same as Figure\,\ref{fig:2306iso}. \label{fig:5288iso}}
\end{figure}

As seen in Figure\,\ref{fig:5288iso}, both stars land on the 2.5\,Gyr stellar isochrone showing that the stars are coeval, as expected. The generated isochrones are evolutionary tracks for stars with metallicity [Fe/H] $= -0.55$, as determined for this system through the DR16 pipeline. The primary star is relatively close to the brown segment of the 2.5\,Gyr isochrone, which denotes the star has almost evolved off the main sequence.

\section{Discussion and Conclusions \label{sec:discuss}}

We thoroughly characterized four benchmark-grade eclipsing binaries, with periods of $\sim \, 3.5$, 9.7, 10.3, and $21$\,d, that have been observed by \kep, \apogee\ and HET. The identified targets were selected from the \kep\ catalog due to their well behaved lightcurves and well sampled radial velocity curves. We further required a significant number ($>$\,8) of well-distributed high-resolution double-lined spectra for the precise determination of stellar masses and the semi-major axis (which provides a scaling factor for the radii). 

For each object we applied \sxcorr, our own implementation of \tdcr, to extract radial velocities. We subsequently generated a binary star model of the light curve and radial velocity data simultaneously using \ph, \emcee\ and \celerite. By incorporating Gaussian processes into the likelihood function, we were able to properly model the stochastic and systematic noise to produce accurate parameter values. 

Finally, we plotted the derived masses and radii against stellar isochrones. This allowed us to validate our assumption of co-eval objects, determine the age of each system and the evolutionary state of each star. Through the use of the stellar isochrone models, we additionally determined that the primary component of KIC\,4076952 and KIC\,5288543 are close to evolving off the main sequence and that the primary component of KIC\,5193386 is a red giant on the red giant branch. These results provide further validation for our semi-automated modeling pipeline through consistency with the \mesa\ {\sc {mist}} isochrones.

\software{
PYTHON \citep{python2009},
KEPHEM \citep{Prsa2011},
PHOEBE \citep{Prsa2005},
EMCEE \citep{DFM2013},
CELERITE \citep{DFM2017},
ASTROPY \citep{astropy2013,astropy2018} }

\section{Acknowledgements}

The authors express their sincere thanks to NASA and the \kep\ team for the amazing, high-quality \kep\ data. The \kep\ mission is funded by NASA's Science Mission Directorate. KH acknowledges support through NASA ADAP grant 80NSSC19K0594. We further acknowledge support from the NSF (grant \#1517592). The authors would additionally like to thank the anonymous referee for their helpful comments that improved this manuscript.

This work was based on observations with the SDSS 2.5-meter telescope. Funding for SDSS-III has been provided by the Alfred P.~Sloan Foundation, the Participating Institutions, the National Science Foundation, and the U.S. Department of Energy of Science. The SDSS-III web site is http://www.sdss3.org/. SDSS-III is managed by the Astrophysical Research Consortium for the Participating Institutions of the SDSS-III Collaboration including the University of Arizona, the Brazilian Participation Group, Brookhaven National Laboratory, University of Cambridge, Carnegie Mellon University, University of Florida, the French Participation Group, the German Participation Group, Harvard University, the Instituto de Astrof\'isica de Canarias, the Michigan State/Notre Dame/JINA Participation Group, Johns Hopkins University, Lawrence Berkeley National Laboratory, Max Planck Institute for Astrophysics, Max Planck Institute for Extraterrestrial Physics, New Mexico State University, New York University, Ohio State University, Pennsylvania State University, University of Portsmouth, Princeton University, the Spanish Participation Group, University of Tokyo, University of Utah, Vanderbilt University, University of Virginia, University of Washington, and Yale University.

Data presented herein were also obtained at the Hobby Eberly Telescope (HET), a joint project of the University of Texas at Austin, the Pennsylvania State University, Stanford University, Ludwig-Maximilians- Universita\"at Mu\"nchen, and Georg-August-Universit\"at G\"ottingen. The HET is named in honor of its principal benefactors, William P. Hobby and Robert E. Eberly.

\bibliography{star_ref}

\end{document}